# High-Quality BN-Graphene-BN Nanoribbon Capacitors Modulated by Graphene Side-gate Electrodes

Yang Wang, Xiaolong Chen, Weiguang Ye, Zefei Wu, Yu Han, Tianyi Han, Yuheng He, Yuan Cai and Ning Wang*

*Department of Physics and the William Mong Institute of Nano Science and Technology, the Hong Kong University of Science and Technology, Hong Kong, China*

Abstract

High-quality BN-Graphene-BN nanoribbon capacitors with double side-gates of graphene are experimentally realized. Graphene electronic properties can be significantly modulated by the double side-gates. The modulation effects are very obvious and followed the metallic electrode behavior of numerical simulations, while the theoretically predicted negative quantum capacitance was not observed, possibility due to the over-estimated or weakened interactions between the graphene nanoribbon and side-gate electrodes.

*Corresponding author: e-mail: phwang@ust.hk

Quantum capacitance (QC) measurement has been proven to be an effective way for probing graphene's density of states (DOS) and electronic properties.[1-3] Different from traditional transport measurement, QC technique is able to directly detect the average DOS of graphene because the measured QC $C_q = e^2\rho$, where $e$ is the electron charge and $\rho$ is the average DOS.[4] Importantly, $C_q$ is more immune to the scattering arising from disorder in graphene samples. Theoretical expressions for the QC of graphene have been verified experimentally.[5-8] Through investigating graphene capacitors, interesting properties such as mid-gap states induced by resonant impurities[9-11] and negative electronic compressibility or negative QC[12,13] have been reported recently. Negative QC has been observed in graphene containing mid-gap states[12,13] or under high magnetic fields.[14] This interesting phenomenon is believed to be related to the electron-electron (e-e) interaction involving fundamental physics of many-body systems.[15-17] Theoretically, negative QC in graphene can also be realized through nanoribbon devices modulated by side-gate electrodes.[18] R. Reiter *et al.* reported numerical simulations of side-gated graphene nanoribbon capacitors and revealed that the capacitive coupling of the nanoribbon and side-gate electrodes should result in negative QC[18]. In this paper, we demonstrate high-quality BN-Graphene-BN nanoribbon capacitors and side-gate modulation effects on graphene properties.

To fabricate high-quality graphene nanoribbon capacitance devices with side-gate electrodes, BN-Graphene-BN capacitors were first prepared using mechanical exfoliation and sample transfer techniques reported previously.[19-25] Fig 1(a) shows schematically the structure of initial capacitance device with graphene side-gate electrodes. A hexagonal boron nitride (h-BN) flake was first exfoliated by the Scotch tape method and placed on the surface of a p-type Si substrate coated with a silicon oxide layer (300 nm thick), serving as the supporting substrate.[22,23] Then a single-layer graphene sheet was exfoliated onto a PMMA membrane which was used to transfer the graphene sheet onto the h-BN flake under an optical microscope.[24,25] By the same method, another h-BN sheet was transferred to cover the graphene sheet, serving as the dielectric medium of the quantum capacitance device. The thicknesses of the two BN sheets are about 10 nm as confirmed by atomic force microscopy (AFM). Finally, the top-gate and drain/source electrodes (Cr/Au=5 nm/40 nm) were fabricated using standard electron-beam (e-beam) lithography and e-beam evaporation.[12, 26]

The capacitance measurements were carried out using a capacitance bridge setup based on previous work.[27] The AC excitation voltage was set to 1 mV in order to achieve high stability and precision simultaneously. The sensitivity of the bridge was ~0.1 fF achieved by placing the high electron mobility transistor and reference capacitor closely to the graphene capacitor. The excitation frequency was 10 kHz, ensuring that the graphene capacitor was fully charged. All wires for the bridge setup were shielded and the p-Si substrate was grounded in order to minimize the parasitic capacitance (< 1 fF). As shown in Fig. 1(b), the total capacitance $C_m$ measured between the top-gate electrode and drain/source electrode is the serial connection of graphene's QC $C_q$ and the capacitance of dielectric layer $C_g$ ($C_m^{-1} = C_q^{-1} + C_g^{-1}$).[5, 6] The voltage applied on the graphene sample (also called the channel voltage $V_{ch}$) is obtained from the charge conservation relation, $V_{ch} = \int_0^{V_{tg}} \left(1 - \frac{C_m}{C_g}\right) dV_{tg}$. It is proportional to the Fermi energy of graphene ($V_{ch} = E_F/e$).[12, 13] Obviously, the plot of $C_q$ vs. $V_{ch}$ reveals the DOS of the graphene sample.

Compared to graphene capacitors made by metal-oxide dielectric materials,[4, 6, 8] the BN-Graphene-BN capacitors showed outstanding performance and clearly revealed the intrinsic properties of graphene with minimal influence of charged impurities or surface disorders at the interface between graphene and the dielectric layer. The high-quality of BN-Graphene-BN capacitors is reflected by the capacitance data shown in Fig. 1(c). The dashed line denotes the capacitance of the BN layer ($C_g = 0.186\ \mu F/cm^2$), which is determined by measuring a reference capacitor made on the same BN layer but without any graphene sheet. The Landau level (LL) oscillations at $N = 0, \pm 1, \pm 2 \ldots$ are clearly observed at a magnetic field of 8 T and the plateaus on these LLs are flat and wide, indicating very few long-range or short-range disorders induced into the graphene capacitance device. Fig. 1(d) illustrates the corresponding $C_q$ as a function of channel voltage $V_{ch}$ in which the LL peaks tend to be infinity, again a clear sign for the high quality of the BN-based graphene capacitor. In addition, the splitting of $0^{th}$ LL and negative QC features are clearly visible in the inset in Fig. 1(c). The 4-fold degeneracy of $0^{th}$ LL (2 for spins and 2 for valleys) is broken into 4 separated levels at 8T.[15, 28] These peaks also exceed the capacitance of BN dielectric layer as denoted by the dashed line, which means the QC of graphene is negative ($C_q^{-1} = C_m^{-1} - C_g^{-1} < 0$).[14, 25] These phenomena could only be observed in an ultra clean 2D sample under a strong magnetic field in which e-e interactions occur.[16] The transport measurements for the BN-Graphene-BN devices determined that the samples showing $0^{th}$ LL splitting [29-32] have a high mobility (178,000 to 195,000 cm$^2$/Vs) at room temperature (see Fig. 2). The transport measurements were conducted using 4-proble configuration separately in order to eliminate the contact resistance. The high quality BN-Graphene-BN structure is important and necessary for fabricating graphene nanoribbon capacitors with side-gate modulations.

To fabricate BN-Graphene-BN nanoribbon capacitors with side-gate electrodes, we designed a zigzag configuration which can effectively increase the length of the nanoribbon and thus its total capacitance. This is because the width of the nanoribbon has to be small enough ($\leq$ 300 nm) in order to get efficient modulations from the side-gate electrodes. On the other hand, the QC of the nanoribbon should be large enough to achieve high precision for the capacitance measurement.[8] Fig. 3(a) is a schematic image of the zigzag graphene capacitor which was realized routinely. Experimentally, similar to the fabrication process mentioned above, the BN-Graphene-BN sandwich-like structure was first prepared. Then the zigzag top-gate electrode, drain/source electrodes and the side-gate electrodes were fabricated by standard e-beam lithography simultaneously.[26] Fig. 3 (c) is the top view of one zigzag capacitance device under an optical microscope. The BN-Graphene-BN sheet was carved along the edges of the top-gate electrode by the Ga$^+$ ion-beam etching in Raith ion LiNE system. The ion beam current was ~5.3 pA at 30 kV. The resolution of the ion beam system is high enough to generate a narrow gap (50 nm) with sharp edges along the cutting line as shown in the scanning electron microscope (SEM) image in Fig. 3(b). The left parts cut by the ion-beam alongside the nanoribbon serve as the graphene side-gate electrodes. The long nanoribbon capacitor is 300 nm in width and 150 μm in length.

Compared to a straight nanoribbon capacitor, the corners of the zigzag configuration were only of small parts causing just minor influence to the whole capacitance measurement. Similar to the pristine graphene capacitor characterization, the measurement of nanoribbon capacitance with side-gate modulations was also carried out on the capacitance bridge system using the same experimental conditions and parameters. The nanoribbon capacitor was first measured with side-gates off. As shown in Fig. 3(d), the measured total capacitance showed a typical V-shape capacitance curve of high-quality graphene sample without introducing serious doping effect since the curve looks symmetrical near the charge neutrality point (CNP) and the CNP is very close to zero energy, demonstrating that the electronic properties and band structures of graphene are not affected after the ion beam cutting process. In addition, compared to the same capacitance measured before cutting, the capacitance per unit area just changed slightly (~0.18 μF/cm²).

The side-gate voltages largely affect the QC measured from the nanoribbon capacitors. We first introduced anti-symmetric voltages into the graphene nanoribbon capacitor, which means the voltages applied on the two side-gate electrodes were opposite ($V_{sg1} = -V_{sg2}$). According to the prediction of previous simulations,[18] the anti-symmetric side-gates can result in an increase of the measured total capacitance and even exceed the capacitance of dielectric layer at the edges of capacitance dip, i.e. negative quantum capacitance ($C_q^{-1} = C_m^{-1} - C_g^{-1} < 0$), mainly due to the capacitive coupling between the graphene nanoribbon and graphene side-gate electrodes. However, the negative QC induced by side-gate modulations was not observed in our experiments, as shown in Fig. 4(a). The capacitance curve became much fatter when the side-gate voltage was increased and the Dirac point was raised up but kept unshifted. In addition, a flat plateau around the Dirac point appeared. The corresponding 2D mapping of measured total capacitance $C_m$ on the $V_{tg} - V_{sg}$ plane was shown in Fig. 4(b), presenting an X-shape evolution. The plateau around Dirac point became thinner and deeper first then recovered fatter and higher as the side-gate voltages increased further. Although the predicted negative quantum capacitance induced by graphene side-gate modulation was not observed, our experimental data were consistent with the numerical simulations of metallic side-gate electrodes (rather than graphene side-gate electrodes) very well.[18] Ideally, this modulation effect is originated from the anti-symmetric voltages between the graphene side-gate electrodes which lead to a non-uniform electric field on the graphene nanoribbon. This is equivalent to the fact that different parts of graphene nanoribbon should reach the Dirac point at different top-gate voltages. On average, the modulated capacitance curve became fat and presented a flat plateau.

Similarly, when symmetric voltages were introduced, which means the voltages applied on the two side-gate electrodes were identical ($V_{sg1} = V_{sg2}$), the modulation effect was also obvious (see Fig. 4(c)). The capacitance curve became fatter and the Dirac point was raised up and shifted to right significantly (for negative side-gate voltages). The corresponding 2D mapping of measured total capacitance $C_m$ on the $V_{tg} - V_{sg}$ plane was shown in Fig. 4(d), presenting a monotonic S-shape evolution. Again, we did not observe the predicted negative quantum capacitance. Our experimental results matched the numerical simulations of metallic side-gate electrodes very well.[18] Why did the modulation effects follow a metallic behavior and the predicted negative quantum capacitance was not detected experimentally? One possible reason was that the capacitive coupling between the graphene nanoribbon and graphene side-gate electrodes might not be as strong as that predicted theoretically. In addition, the gaps and width of graphene nanoribbon were a little larger than the theoretical ones, and this might also weaken the interactions between the graphene side-gate electrodes and nanoribbon. Nevertheless, the side-gate voltages indeed have significantly modulated the graphene electronic properties, no matter for symmetric or anti-symmetric side-gate voltages. The present nanoribbon capacitors with double side-gate structure demonstrated a possible nano-device structure for potential technological applications.

**Acknowledgement**

The authors are grateful to Prof. H.B. Chan for his great help on the capacitance bridge setup for this research. Financial supports from the Research Grants Council of Hong Kong (Project Nos. 604112, N_HKUST613/12 and HKUST9/CRF/08, HKUST/SRFI) and technical support of the Raith-HKUST Nanotechnology Laboratory for the electron-beam lithography facility (Project No. SEG HKUST08) are hereby acknowledged.

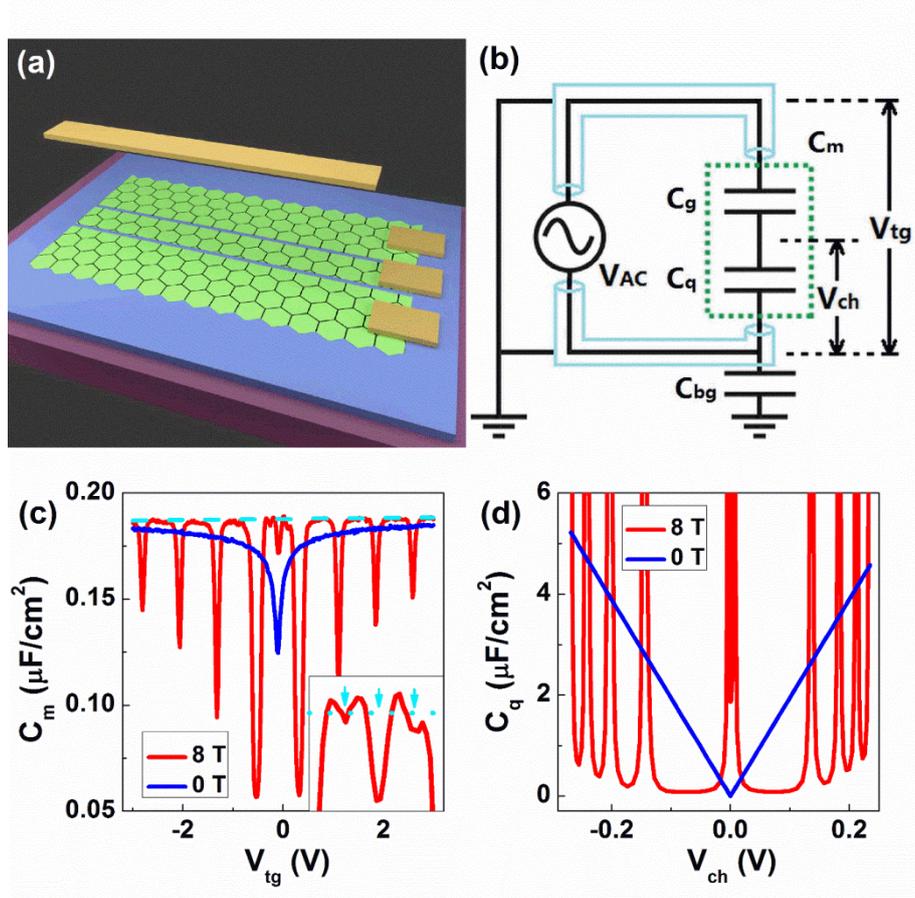

Fig. 1 (a) Schematic image of BN-Graphene-BN capacitance device with graphene side-gate electrodes, where the top BN sheet is hidden. The purple plate denotes the silicon oxide substrate and the sapphire sheet is the bottom BN sheet. The top-gate, drain/source and side-gate electrodes are colored by gold. (b) The equivalent circuit of quantum capacitance measurement. (c) Measured total capacitance $C_m$ as a function of top-gate voltage $V_{tg}$ under $B = 0\ \text{T}$ (blue line) and $8\ \text{T}$ (red line). The cyan dashed line denotes the capacitance of dielectric medium $C_g = 0.186\ \mu\text{F/cm}^2$. The inset shows the zoom-in details around the 0$^{\text{th}}$ LL. The cyan arrows denotes the LL splitting and the peaks exceeding the cyan dotted line reveal the negative quantum capacitance. (d) Corresponding quantum capacitance $C_q$ as a function of channel voltage $V_{ch}$ under $8\ \text{T}$ (red line). The blue line denotes the theoretical value under $0\ \text{T}$.

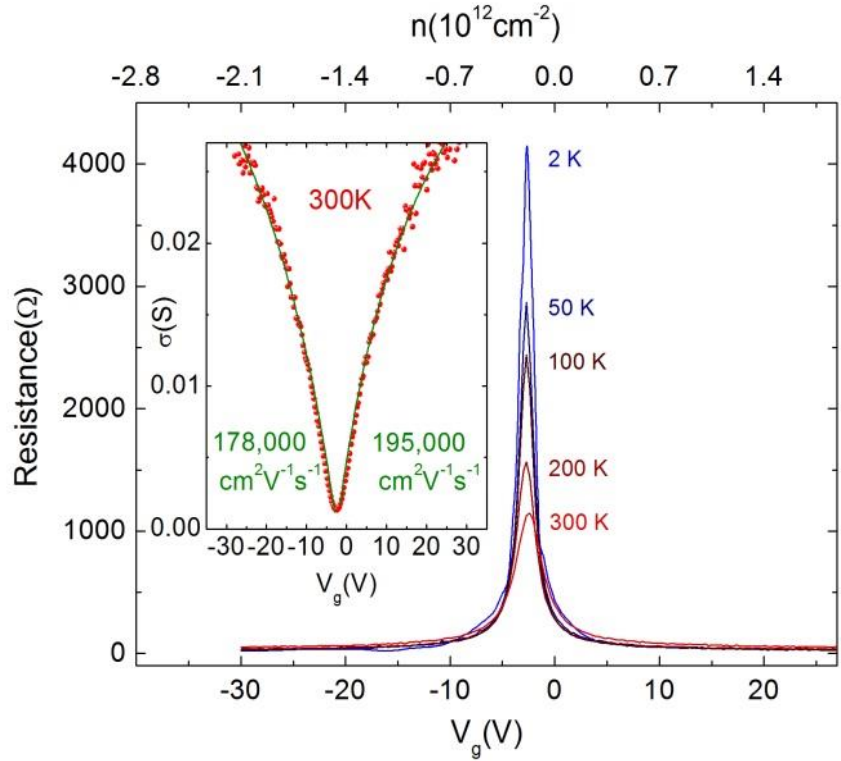

Fig. 2 Transport data of the BN-graphene-BN sample measured using four-probe configuration at different temperatures. Inset image showed a high mobility ~ 195,000 cm$^2$V$^{-1}$s$^{-1}$ of the sample at room temperature. The mobility of graphene is fitted with the formula $\sigma^{-1} = (\mu n e)^{-1} + \sigma_s^{-1}$ (green solid line), where μ is the mobility and $\sigma_s$ is the conductivity due to short-range scatterings.

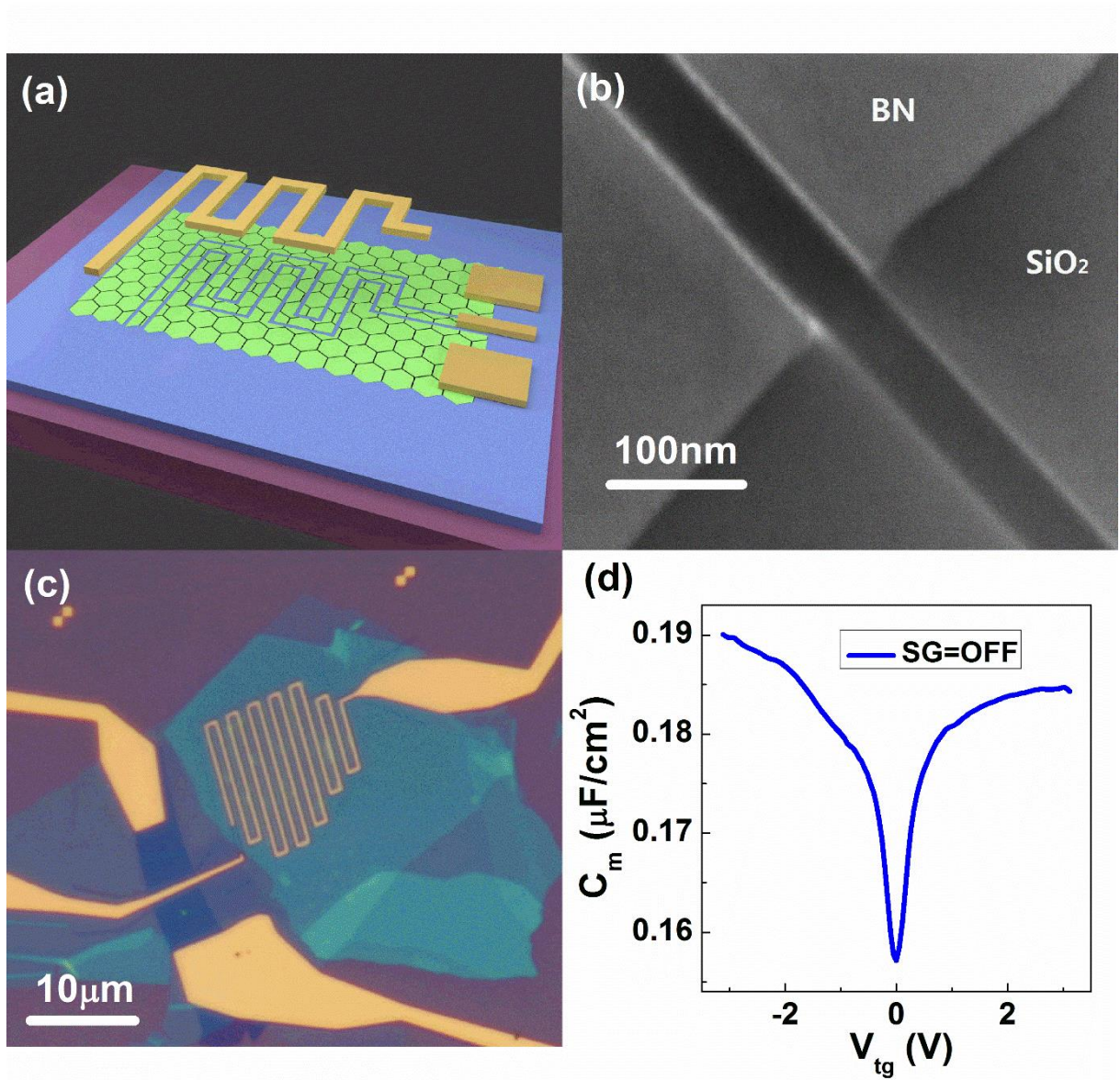

Fig. 3 (a) Schematic image of the graphene nanoribbon capacitor with side-gate modulations. The graphene sheet (green membrane) is carved into zigzag configuration with left parts serving as the side-gate electrodes. (b) SEM image of the cutting line by the Ga$^+$ ion-beam technique. The width of the gap is about 50 nm. (c) Optical image of the fabricated capacitance device (300 nm in width and 150 μm in length. (d) Measured total capacitance $C_m$ as a function of top-gate voltage $V_{tg}$ at 300 K with side-gates off.

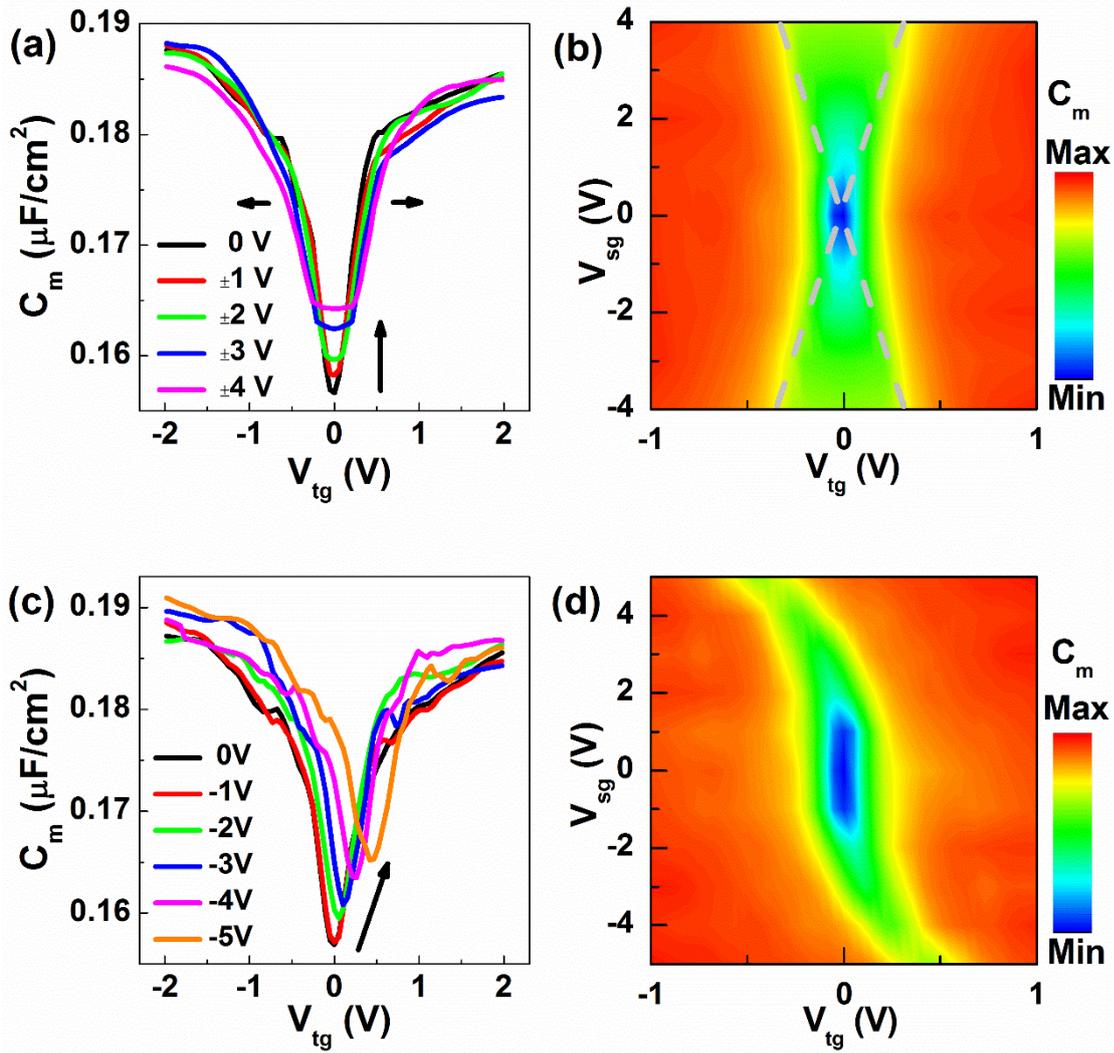

Fig. 4 (a) Measured total capacitance $C_m$ as a function of top-gate voltage $V_{tg}$ for different anti-symmetric side-gate voltages ($V_{sg1} = -V_{sg2} = 0\,\text{V}, 1\,\text{V} \ldots 4\,\text{V}$). The arrows denote the modulation effect on capacitance curves. (b) Corresponding 2D mapping of measured total capacitance $C_m$ on the $V_{tg} - V_{sg}$ plane, presenting an X-shape evolution (gray dashed lines). (c) Measured total capacitance $C_m$ as a function of top-gate voltage $V_{tg}$ for different symmetric side-gate voltages ($V_{sg1} = V_{sg2} = 0\,\text{V}, -1\,\text{V} \ldots -5\,\text{V}$). The arrows denote the modulation effect on capacitance curves. (d) Corresponding 2D mapping of measured total capacitance $C_m$ on the $V_{tg} - V_{sg}$ plane, presenting an S-shape evolution.